\documentstyle[prl,aps]{revtex}
\begin{document}
\draft
\title{Dirac and Weyl Equations on a Lattice as Quantum Cellular
Automata}
\author{Iwo Bialynicki-Birula}
\address{Centrum Fizyki Teoretycznej PAN$\,^*$\\
Lotnik\'ow 32/46, 02-668 Warsaw, Poland\\ and \\
Institut f\"ur Theoretische Physik, Johann Wolfgang
Goethe-Universit\"at\\ Robert-Mayer-Strasse 8-10, Frankfurt am
Main, Germany}

\maketitle
\begin{abstract}
A discretized time evolution of the wave function for a Dirac
particle on a cubic lattice is represented by a very simple quantum
cellular automaton. In each evolution step the updated value of the
wave function at a given site depends only on the values at the
nearest sites, the evolution is unitary and preserves chiral
symmetry. Moreover, it is shown that the relationship between Dirac
particles and cellular automata operating on two component objects
on a lattice is indeed very close. Every local and unitary
automaton on a cubic lattice, under some natural assumptions, leads
in the continuum limit to the Weyl equation. The sum over histories
is evaluated and its connection with path integrals and theories of
fermions on a lattice is outlined.
\end{abstract}


The aim of this paper is to present a very simple lattice algorithm
that reproduces in the continuum limit the propagation of massless
or massive Dirac particles. The discretized time evolution
described here satisfies the fundamental physical requirements of
locality (the updated value of the wave function at a given site
depends only on the values at the neighboring sites), unitarity
(the norm of the wave function is preserved), and chiral symmetry
(independent propagation of two helicity states).

The main, and totally unexpected, result of this investigation is
a discovery that for two-component wave functions on a cubic
lattice the Weyl equation necessarily follows in the continuum
limit from locality, unitarity, and two additional, natural
assumptions: (A) wave functions that have constant values
throughout the lattice do not change in time and (B) the evolution
algorithm preserves the symmetry of the lattice. Thus, the rotation
group and spin emerge automatically in the continuum limit from
unitary dynamics on a cubic lattice.

The results presented in this paper are directly related to
numerous proposals of path integrals for a Dirac particle since an
iteration of a discretized time evolution gives a sum over
histories. The path integral for the Dirac particle in one space
dimension was found by Feynman \cite{f,fh} and independently by
Riazanov \cite{ria}. Even this relatively simple problem, where
there is no spin to complicate matters, is still attracting
attention \cite{tn,ko}. There is no consensus at all as to the form
of path integrals for a Dirac particle in three dimensions.
Previously proposed path integrals fall into three broad
categories. The first category
\cite{ria,pd,ros,ki,gjks,bd,it,pol,adj} comprises those approaches
that work by ``reverse engineering'' introducing from the outset
the Dirac matrices to describe the spin degrees of freedom. These
approaches are often combined with the proper-time representation
of the Dirac operator. Into the second category \cite{sch,tj1,tj2}
fall those formulations that derive spin from continuously
parametrized space of states related to the rotation group. To the
third category \cite{bm,rav,os,gs} belong all approaches based on
anticommuting Grassmannian variables. The sum over histories that
is obtained from my lattice algorithm is distinct from all these
path integrals since the evolution is fully discretized and the
rotation group emerges only in the continuum limit.

There is an obvious connection between this work and theories of
fermions on a lattice but there is also an essential difference in
the choice of objects to be calculated. I make no attempt to write
down discretized versions of the Hamiltonian or the Lagrangian
working instead with a discretized version of the evolution
operator. In my approach common difficulties (breaking of chiral
symmetry, doubling of fermion species, nonlocality) encountered in
formulating the dynamics of a Dirac particle on a lattice
\cite{kw,ks,ls,dwy,rab,bms,kt} and the no-go theorem concerning
Weyl particles on a lattice \cite{nm} are avoided.

Algorithms for cellular automata describing quantum system have
been studied by Groessing and Zeilinger \cite{gz}, but their
discretized time evolution does not conserve probability. In a
recent paper Kostin \cite{kos} has introduced a cellular automaton
for the Dirac equation that does conserve probability but his
algorithm is nonlinear and it gives a linear evolution equation
only in the continuum limit. Following the example set by these
authors I use the term {\it quantum cellular automaton} to denote
discretized evolution of complex wave functions. I believe that the
name {\it cellular automaton} is justified despite the fact that
one of the eight properties of cellular automata required by
Wolfram \cite{wol} does not hold: the states are described by
continuous, not by discrete variables. However, my algorithm for
updating the state of the system is synchronous, homogeneous,
discrete in space and in time, deterministic, and spatially and
temporally local. In my case the name {\it quantum} is fully
justified by the unitarity of time evolution.

Let me begin with a massless particle on a cubic lattice. In my
quantum cellular automaton the two-component wave function
$\phi(i,j,k,t)$ is updated for each time increment $\Delta t$
according to the following local algorithm
\begin{eqnarray}
\phi(i,j,k,t+\Delta t) = W_{+++}\phi(i+1,j+1,k+1,t)\nonumber\\
+ W_{++-}\phi(i+1,j+1,k-1,t)+\cdots \nonumber\\
 + W_{---}\phi(i-1,j-1,k-1,t),\label{a}
\end{eqnarray}
where the integers $i,j$ and $k$ are the coordinates of lattice
sites in units of the cell size $a$ (${\bf r} = (a i,a j,a k)$) and
all eight $W$'s are $2\times 2$ matrices. This algorithm can be
written in a compact form

\begin{eqnarray}
\phi({\bf r},t + \Delta t) = \sum_{\bf h} W({\bf h})\phi({\bf r} +
{\bf h},t),\label{a1}
\end{eqnarray}
where the summation extends over a set of vectors ${\bf h}$ that
point from a given site to eight nearest sites. Their components in
units of $a$ are $\pm 1$,

\begin{eqnarray}
\{{\bf h}\} =
\{\{1,1,1\},\{1,1,-1\},\{1,-1,1\},\{1,-1,-1\}\nonumber\\
\{-1,1,1\},\{-1,1,-1\},\{-1,-1,1\},\{-1,-1,-1\}\}.\label{set}
\end{eqnarray}

Upon evaluating the norm of the updated wave function, one finds
that the matrices $W({\bf h})$ must satisfy the following algebraic
relations to guarantee the unitarity of the transformation
(\ref{a1})

\begin{eqnarray}
\sum_{\bf h}W^{\dag}({\bf h})W({\bf h}) = 1,\label{c1}\\
\sum_{\bf h}W^{\dag}({\bf h})W({\bf h+h'-h''}) = 0,\label{c2}
\end{eqnarray}
where ${\bf h'}$ and ${\bf h''\neq h'}$ are arbitrary vectors
belonging to the set (\ref{set}) and the sum in (\ref{c2}) extends
only over vectors ${\bf h}$ such that the vectors ${\bf h+h'-h''}$
are also members of the set. It follows from these unitarity
conditions that the inverse of (\ref{a1}) has a similar local form
so that my automaton is fully reversible,

\begin{eqnarray}
\phi({\bf r},t) = \sum_{\bf h} W^{\dag}({\bf h})\phi({\bf r} + {\bf
h},t+\Delta t).\label{a2}
\end{eqnarray}

Since the inverse of a unitary transformation is also unitary, the
hermitian conjugate matrices $W^{\dag}$ must obey the same
conditions as the matrices $W$,

\begin{eqnarray}
\sum_{\bf h}W({\bf h})W^{\dag}({\bf h}) = 1,\label{c3}\\
\sum_{\bf h}W({\bf h})W^{\dag}({\bf h+h'-h''}) = 0.\label{c4}
\end{eqnarray}

Depending on the mutual orientation of the vectors ${\bf h'}$ and
${\bf h''}$, the conditions (\ref{c2}) and (\ref{c4}) have one,
two, or four terms, as exemplified below

\begin{mathletters}
\begin{eqnarray}
W^{\dag}_{+++}W_{---} = 0,\label{u3a}\\
W^{\dag}_{+++}W_{+--} + W^{\dag}_{-++}W_{---} = 0,\label{u3b}\\
W^{\dag}_{+++}W_{++-} + W^{\dag}_{+-+}W_{+--}
+ W^{\dag}_{-++}W_{-+-}\nonumber\\ + W^{\dag}_{--+}W_{---} =
0.\label{u3c}
\end{eqnarray}
\end{mathletters}
In total there are 8 conditions of the type (\ref{u3a}), 12
conditions of the type (\ref{u3b}), 6 conditions of the type
(\ref{u3c}), and one condition (\ref{c1}) and then the same number
of conditions with $W$ and $W^{\dag}$ interchanged. Nonetheless
they can all be satisfied by the following simple and symmetric
choice of $W$'s

\begin{eqnarray}
W_{+++}&=&q_+P_1,\,\,W_{++-}=q_-P_2,\,\,W_{+-+}=q_-P_1,\nonumber\\
W_{+--}&=&q_+P_2,\,\,W_{-++}=q_-P_3,\,\,W_{-+-}=q_+P_4,\nonumber\\
W_{--+}&=&q_+P_3,\,\,W_{---}=q_-P_4,\label{w3}
\end{eqnarray}
where

\begin{eqnarray}
q_+ = (1+i)/4,\,\,\,q_- = (1-i)/4,
\end{eqnarray}
and

\begin{eqnarray}
P_1 &=& \left(\begin{array}{cc}
\,1 & \,\,0\\
\,1 & \,\,0\end{array}\right),\,\,\,
P_2 = \left(\begin{array}{cc}
0 & \,\,1\\
0 & \,\,1\end{array}\right),\nonumber\\
P_3 &=& \left(\begin{array}{cc}
1 & 0\\
-1 & 0\end{array}\right),\,\,
P_4 = \left(\begin{array}{cc}
0 & -1\\
0 & 1\end{array}\right).\label{p}
\end{eqnarray}

In order to find the continuum limit of the evolution equation I
shall use the exponential representation of the shift operators to
write (\ref{a1}) in the form

\begin{eqnarray}
\phi({\bf r},t + \Delta t) = \sum_{\bf h} W({\bf h}) \exp({\bf
h}\cdot{\nabla})\phi({\bf r},t).\label{a3}
\end{eqnarray}
Expanding both sides of this equation in powers of $\Delta t$ and
$a$ (${\bf h}$ is of the order of $a$) and keeping only linear
terms, I obtain the Weyl equation,

\begin{eqnarray}
\partial_t \phi({\bf r},t) = c\,\mbox{\boldmath
$\sigma$}\!\cdot\!\nabla\phi({\bf r},t),
\end{eqnarray}
where $c = a/\Delta t$. Incidentally, to obtain a discretized Weyl
equation in two space dimensions one may choose the four matrices
$W_{++},W_{+-},W_{-+}$, and $W_{--}$ as equal to the matrices
$P_i$.

I shall show now that {\it for every set} of $2\times 2$ matrices
satisfying the unitarity conditions, and not just for the
particular choice (\ref{w3}), one obtains the Weyl equation in the
continuum limit. To prove this let me first expand the exponential
operator in (\ref{a3}) into powers of $a$,

\begin{eqnarray}
\sum_{\bf h} W({\bf h}) \exp({\bf h}\cdot{\nabla}) = 1 + a\,{\bf
s}\!\cdot\!\nabla + \cdots, \label{exp}
\end{eqnarray}
where I have made use of the assumption A that a homogeneous wave
function does not change in time,

\begin{eqnarray}
\sum_{\bf h} W({\bf h}) = 1,\label{sum}
\end{eqnarray}
and I have introduced the matrices $s_i$,

\begin{mathletters}
\label{ss}
\begin{eqnarray}
s_x = W_{+++}+W_{++-}+W_{+-+}+W_{+--}\nonumber\\
-W_{-++}-W_{-+-}-W_{--+}-W_{---},\\
s_y = W_{+++}+W_{++-}-W_{+-+}-W_{+--}\nonumber\\
+W_{-++}+W_{-+-}-W_{--+}-W_{---},\\
s_z = W_{+++}-W_{++-}+W_{+-+}-W_{+--}\nonumber\\
+W_{-++}-W_{-+-}+W_{--+}-W_{---}.
\end{eqnarray}
\end{mathletters}

Since the differentiations are antihermitian operators, the
unitarity of the operator (\ref{exp}) requires that the three
matrices $s_i$ must be hermitian and, therefore, the formulas
(\ref{ss}) must also hold with all the matrices $W$ replaced by
their hermitian conjugates. This fact will enable me to use the
unitarity conditions (\ref{c2}) and (\ref{c4}) to calculate
products of the matrices $s_i, (i=x,y,z)$. The unitarity conditions
also imply (to see this, one has to write them down explicitly)
that all products of a $W$ matrix by its hermitian conjugate
commute, and therefore they can be simultaneously diagonalized. From
the equivalence of all six lattice directions (assumption B)
I conclude that the eigenvalues of all matrices $W^{\dag}({\bf
h})W({\bf h})$ must be the same and then from equations (\ref{c1})
and (\ref{c3}) I find that these eigenvalues are $1/4$ and $0$. It
is now a matter of tedious but otherwise straightforward algebraic
manipulations to show that the matrices $s_i$ satisfy the familiar
anticommutation relations,

\begin{eqnarray}
s_i s_j + s_j s_i = 2\delta_{ij}.\label{pau}
\end{eqnarray}
There exist only two inequivalent two-dimensional representations
of these relations: \,${\bf s} = \mbox{\boldmath $\sigma$}$ or
${\bf s} = -\mbox{\boldmath $\sigma$}$. They describe the
propagation of two helicities. Thus, up to a choice of helicity,
the universality of the Weyl equation under the listed assumptions
is established.

For massless particles each helicity state propagates
independently. Chiral invariance (or CP symmetry) is expressed in
my discretized form of time evolution by the fact that the matrices
$W$ corresponding to the two helicities are related by the spatial
reflection

\begin{eqnarray}
W({\bf h})\, \to\, W({-\bf h}),\label{par}
\end{eqnarray}
as seen from the formulas (\ref{ss}). For massive particles,
however, the two helicity states are mixed by the mass term.
Therefore, the discretized time evolution for a massive particle
must be described, as in the standard Dirac equation, in terms of
two two-component wave functions. The discrete time-evolution
algorithm for a massive Dirac particle can again be written in the
same general form (\ref{a2}) as for the massless case

\begin{eqnarray}
\psi({\bf r},t + \Delta t) = \sum_{\bf h} D({\bf h})\psi({\bf r} +
{\bf h},t).\label{a4}
\end{eqnarray}
The $4\times 4$ matrices $D$ that act on four-component wave
functions $\psi({\bf r},t)$ can be expressed in terms of the
matrices $W({\bf h})$ and $W({-\bf h})$ by the following
block-matrix formulas (from now on I set $c=\Delta t/a = 1$ and
$\hbar = 1$)

\begin{eqnarray}
D({\bf h}) = \left(\begin{array}{cc}
\cos(m a) & \!i\sin(m a)\\
\!i\sin(m a) & \cos(m a)\end{array}\right)
\!\!\left(\begin{array}{cc}
\!W({\bf h}) & \,0\\
\,0 & \!\!W({-\bf h}) \end{array}\right).
\end{eqnarray}
One may check that such a modification does not affect the
unitarity conditions; the matrices $D$ will also satisfy all of
them. The continuum limit of (\ref{a4}) gives the Dirac equation in
the Weyl representation of the Dirac matrices

\begin{eqnarray}
\partial_t \psi({\bf r},t) = (\rho_3\,\mbox{\boldmath
$\sigma$}\!\cdot\!\nabla + i\rho_1m)\psi({\bf r},t).
\end{eqnarray}

I shall now write down the sum over histories for the Weyl particle
arising from my time-evolution algorithm (\ref{a1}) and I shall
directly show that it yields the propagator in the continuum limit.
The $N$-fold iteration of the single-step evolution leads to the
formula

\begin{eqnarray}
&&\phi({\bf r},t + N\Delta t)\nonumber\\
&& = \sum_{{\bf h}_1,\dots,{\bf h}_N} \!\!\!W({{\bf h}_1})\cdots
W({{\bf h}_N})\phi({\bf r} + {\bf h}_1 + \cdots + {\bf
h}_N,t).\label{sumh}
\end{eqnarray}
When the sum in this formula is restricted to only those
combinations of the vectors ${\bf h}_i$ that produce a given
displacement, we will obtain a discrete version of the propagator
\begin{eqnarray}
K({\bf r-r'},t) = \!\!\!\!\!\sum_{{\bf h}_1,\dots,{\bf h}_N}
\!\!\!\!\!W({{\bf h}_1})\!\cdots \!W({{\bf h}_N})\delta_{{\bf
r},{\bf r}'+ {\bf h}_1 + \cdots + {\bf h}_N},\label{sump}
\end{eqnarray}
where $t=N\Delta t$. Each term in this sum corresponds to a lattice
trajectory that in $N$ steps connects the initial and the final
lattice sites. A single step described by a vector ${\bf h}$ is
represented by the matrix $W({\bf h})$. Contributions from all
trajectories are coherently added.

The constraint on the sum in (\ref{sump}) can be handled (cf., for
example, \cite{js}) with the help of the Fourier representation of
the Kronecker delta, leading to the following integral form of the
propagator

\begin{eqnarray}
K({\bf r-r'},t) = (\frac{a}{2\pi})^3\!\!\int\!\! d^3k\,\tilde
W^{\frac{t}{\Delta t}}\!({\bf k})\exp[i{\bf k}\!\cdot\!({\bf
r-r'})],\label{prop}
\end{eqnarray}
where

\begin{eqnarray}
\tilde W({\bf k}) = \sum_{\bf h}W({\bf h}) \exp(i{\bf
k}\!\cdot\!{\bf h}),\label{wk}
\end{eqnarray}
and the integration extends over the Brillouin zone:
$\,-\pi/a<k_i<\pi/a$.

The integral (\ref{prop}) can be evaluated with the use of the
explicit representation (\ref{w3}) of the matrices $W$,

\begin{eqnarray}
\tilde W({\bf k}) = c_x c_y c_z + s_x s_y s_z + i[(s_x c_y c_z -
c_x s_y s_z)\sigma_x\nonumber\\
+ (c_x s_y c_z + s_x c_y s_z) \sigma_y + (c_x c_y s_z - s_x s_y
c_z)\sigma_z],\label{wk1}
\end{eqnarray}
where

\begin{eqnarray}
c_i = \cos(k_i a),\,\,\,s_i = \sin(k_i a).
\end{eqnarray}
This is a unitary matrix and its eigenvalues $\lambda_{\pm}$ are

\begin{eqnarray}
\lambda_{\pm} &=& \exp(\pm i\varphi)\nonumber\\
&=& c_x c_y c_z + s_x s_y s_z \pm i\sqrt{1-(c_x c_y c_z + s_x s_y
s_z)^2}.\label{eig}
\end{eqnarray}
An easy way to evaluate the $N$-th power ($N = t/\Delta t$) of the
matrix (\ref{wk1}), needed to calculate the propagator, is to write
$\tilde W$ in terms of its eigenvalues and its unitary
diagonalizing matrix $V$,

\begin{eqnarray}
\tilde W({\bf k}) = V\left(\begin{array}{cc}
\exp(i\varphi) & \,\,0\\
\,\,0 & \exp(-i\varphi) \end{array}\right) V^{\dag},
\end{eqnarray}

\begin{eqnarray}
(\tilde W({\bf k}))^{\frac{t}{\Delta t}} = V\left(\begin{array}{cc}
\exp(it\varphi/a) & \,\,0\\
\,\,0 & \exp(-it\varphi/a) \end{array}\right) V^{\dag}.
\end{eqnarray}
In the limit, when $a\to 0$, one obtains

\begin{eqnarray}
\lim_{a\to 0} (\varphi/a) &=& \vert {\bf k}\vert =: k,\\
\lim_{a\to 0}V &=& ({\bf k}\!\cdot\!\mbox{\boldmath
$\sigma$}-k\sigma_z)/\sqrt{2k(k-k_z)},
\end{eqnarray}
and finally

\begin{eqnarray}
\lim_{\Delta t \to 0}(\tilde W({\bf k}))^{\frac{t}{\Delta t}} =
\exp(i{\bf k}\!\cdot\!\mbox{\boldmath $\sigma$}t),
\end{eqnarray}
in accordance with the Weyl equation. Thus, the sum over histories
(\ref{sumh}) reproduces correctly the propagator in the continuum
limit.

Having reproduced the propagation of a massless particle I can
easily include the mass in the sum over histories since that part
has been already solved by Feynman \cite{f,fh} in one dimension.
More recently Feynman's "checkerboard" picture of a particle
zigzagging through spacetime, reversing its helicity at each bend,
has also been described by a Poisson process \cite{gjks,tj1,gs,js}.
This Poisson process must be combined with the propagation of
definite helicity states. Therefore, the propagator for a massive
Dirac particle will be a sum of terms, each term describing a fixed
number of helicity reversals. Between the reversals the propagation
is described by the sum over histories (\ref{sump}) separately for
each helicity.

Interaction with the electromagnetic field may be accounted for in
the standard fashion by changing the phase of the wave function at
each propagation step according to the local value of the
electromagnetic potential.

I would like to thank Jerzy Kijowski and Joachim Reinhardt for
discussions and Walter Greiner for his hospitality at the
University of Frankfurt.

\end{document}